\documentclass
[12pt]{article}
\title{{\bf Calculation of gluon and four-quark condensates
from the operator expansion}}
\author{B.V.Geshkenbein\thanks{E-mail: geshken@heron.itep.ru}\\
Institute of Theoretical and Experimental Physics,\\
B.Cheremushkinskaya 25, 117218 Moscow,Russia}
\date{}
\begin{document}

\maketitle

\newcommand{\be}{\begin{equation}}
\newcommand{\ee}{\end{equation}}

\def\la{\mathrel{\mathpalette\fun <}}
\def\ga{\mathrel{\mathpalette\fun >}}
\def\fun#1#2{\lower3.6pt\vbox{\baselineskip0pt\lineskip.9pt
\ialign{$\mathsurround=0pt#1\hfil##\hfil$\crcr#2\crcr\sim\crcr}}}

The magnitudes of gluon and four-quark condensates are found from
the analysis of vector mesons consisting of light quarks (the
families of $\rho$ and $\omega$ mesons) in the 3 loops
approximation. The QCD model with infinite number of vector mesons
is used to describe the function $R(s)$. This model describes well
the experimental function $R(s)$. Polarization operators
calculated with this model coincide with the Wilson operator
expansion at large $Q^2$. The improved perturbative theory, such
that the polarization operators have correct analytical
properties, is used. The result is $\langle 0 \vert (\alpha_s/\pi)
G^2 \vert 0 \rangle = 0.062 \pm 0.019 GeV^4$. The electronic
widths of $\rho(1450)$ and $\omega(1420)$ are calculated.

\newpage

{\bf I.~INTRODUCTION}

\vspace{5mm}

The purpose of this work is to propose a new method of calculation
of gluon and other condensates in some loop approximation from
analysis of families of vector mesons consisting of light quarks
($\rho$, $\omega$-families).Calculation of gluon and other
condensates is based on the Wilson operator expansion (OPE)
(formulae (7-10)) for polarization operator
$(\Pi^{(\rho)}(Q^2))_{theor}$. On the other hand, the dispersion
relation for $((\Pi^{(\rho)}(Q^2))_{exp}$ (Eq.3) makes it possible
to express $(\Pi^{(\rho)}(Q^2))_{Exp}$ through the measurable
function $R^{(\rho)}(s)$. At large $Q^2$~
$(\Pi^{(\rho)}(Q^2))_{Exp}$ must coincide with
$(\Pi^{(\rho)}(Q^2))_{theor}$. In order to
present$(\Pi^{(\rho)}(Q^2))_{Exp}$ in the form (7) we use the QCD
model with an infinite number of vector mesons (MINVM) suggested
and used in papers [1-3].  The MINVM was used in papers [4,5] for
calculation of hadronic contribution to muon ($g-2)$-factor and
$\alpha(M^2_Z)$. The accuracy of these calculations by MINVM is
about 1\%. It is evident that the MINVM can be used only under the
 integral.

This work is the continuation of Ref.[6] devoted to evaluating  of
the QCD parameters. In Ref.[6] the analiticity  of QCD
polarization operators were combined with the renormalization
group and was used for investigation of hadronic $\tau$-decay.

The calculation according to renormalization group leads to
appearance of nonphysical singularities. So, the one-loop
calculation gives a nonphysical pole, while in the calculation in
a larger number of loops the pole disappears, but a nonphysical
cut appears $[-\Lambda^2_3, 0]$.\footnote{This is the definition
of $\Lambda_3$.} In Ref.[6] it was shown that there are only two
values of $\Lambda_3$, such that theoretical predictions of QCD
for $R_{\tau, V+A}$ (formulae (23) and (24) of [6]) agree with the
experiments [7-9]. These values calculating in the three loops are
the following: one conventional value $\Lambda^{conv}_3= (618\pm
29)MeV$ and the other value of $\Lambda_3$ is $\Lambda^{new}_3 =
(1666\pm 7) MeV$. The predictions of QCD consistent with the
experiments [7-9] are just within these values of $\Lambda_3$. If
one simply takes off the nonphysical cut and leaves the
conventional value $\Lambda^{conv}_3$ then discrepancy between the
theory and experiment will arise. As was shown in [6], if instead
of conventional value $\Lambda^{conv}_3$ one chooses the value
$\Lambda^{new}_3 = (1565 \pm 193)MeV$ , then only the physical cut
contribution is sufficient to explain the experiment on hadronic
$\tau$-decay.

The new sum rules following only from analytical properties of the
polarization operator were obtained in [6].These sum rules imply
that there is an essential discrepancy between perturbation theory
in QCD and the experiment in hadronic $\tau$-decay at conventional
value of $\Lambda_3$. If $\Lambda_3 = \Lambda^{new}_3$, this
discrepancy is absent. Because $\Lambda^{conv}_3$ is conventional,
we will calculate the condensates for both admissible values of
$\Lambda_3$; $\Lambda^{new}_3 = (1565 \pm 193) MeV$ (in three
loops)  without the nonphysical cut corresponding
$\alpha_s(-m^2_r)=0.379\pm 0.013$ and $\Lambda^{conv}_3 = (618 \pm
29) MeV$ with the nonphysical cut corresponding
$\alpha_s(-m^2_r)=0.354\pm 0.010$.

The paper is organized as follows. In Sec.II the QCD model with an
infinite number is expanded. The polarization operator defined
from experiment with the help of formulae (3-6) has the form of
OPE and  owing to this, it is needless to use the Borel
transformation. To find the condensates it is sufficient to equate
the coefficients at $1/Q^n$. In Sec.III the magnitudes of gluon
and four-quark condensates are calculated from analysis of
$\rho$-meson family without taking into account  $\rho-\omega$
interference. It is obtained $\langle 0\vert (\alpha_s/\pi) G^2
\vert 0 \rangle = (0.074 \pm 0.023) GeV^4$ $(\Lambda_3 = 1.565
GeV)$ and $\langle 0 \vert (\alpha_s)/\pi G^2 \vert 0 \rangle =
(0.112 \pm 0.021) GeV^4$~ ($\Lambda_3 = 0.618 GeV$).

In Sec.IV the magnitudes of gluon and four-quark condensates are
calculated from analysis of $\omega$-meson family without taking
into account $\rho-\omega$ interference. It is obtained $\langle 0
\vert (\alpha_s/\pi) G^2 \vert 0 \rangle = (-0.076 \pm 0.033)
GeV^4$ ~ ($\Lambda_3 = 1.565 GeV$) and $\langle 0 \vert
(\alpha_s/\pi) G^2 \vert 0 \rangle = (-0.043 \pm 0.031)GeV^4$~
$(\Lambda_3 = 0.618 GeV^4$).

The magnitude of gluon condensate obtained from analysis of
$\omega$-family must be equal to the magnitude of gluon condensate
from analysis of $\rho$-family.

This discrepancy is due to that all formulae obtained from OPE are
valid only for states with pure isospin. But, owing to the
vicinity of the mass of the $\rho$ and $\omega$-mesons the
$\rho$-meson has a small admixture  of the state with isospin
$I=0$ and the $\omega$-meson has a small admixture of the state
isospin $I=1$. The contradiction  resolves after the separation
from the $\rho$ and $\omega$ mesons of the pure states $\rho_0$
with the isospin $I=1$ and $\omega_0$ with isospin $I=0$. It must
be emphasized that strong cancellations occur in the formulae
determining the condensates. For these reasons the account  of the
fine effects as the $\rho-\omega$ interference is essential. In
Sec.V the analysis of $\omega$-family with account of the
$\rho-\omega$ interference is given. The results of the
calculations in the ($0-3)$loop approximation of gluon  and
four-quarks condensates are presented in Table I ($\Lambda_3=1.565
GeV)$ and in Table II ($\Lambda_3=0.618 GeV)$. The same analysis
of $\rho$-family is given in Sec.VI (Table III,IV). Because  the
values of the condensates obtained from the analysis of the $\rho$
and $\omega$ families agree closely in Sec.IX  the averaged values
of the condensates (Table V,VI) are presented. From these tables
it is evident that the expansion of the gluon and 4-quarks
condensates in terms of $\alpha_s$ is very good. As a by product
in Sec's.VII and VIII, we calculated the electronic widths of
$\rho(1450)$ and $\omega(1420)$.

\vspace{7mm}

\centerline{\bf II.~ THE QCD MODEL WITH AN  INFINITE NUMBER}

\centerline{OF VECTOR  MESONS.}

\vspace{3mm}
 In this Section we present the QCD model with infinite number of
 vector mesons suggested in Refs.[1-3]. This model is a basis for
 the following calculations.

 Let us consider at first the family of $\rho$-mesons. The polarization
 operator $\Pi^{(\rho)}$ corresponding to the $\rho$-meson family
 has the form
 \be
 i \int d^4 x e^{iqx} \langle 0 \vert T \{j^{I=1}_{\mu} (x),
 j^{I=1}_{\nu} (0)\} \vert 0 \rangle = (q_{\mu} q_{\nu} - q^2 g_{\mu
 \nu}) \Pi^{(\rho)} (Q^2),
 \ee
 where $Q^2 = -q^2$ and
 \be
 j^{I=1}_{\mu} (x) = (1/2) [\bar{u} (x) \gamma_{\mu} u(x) - \bar{d} (x)
 \gamma_{\mu} d(x) ]
 \ee
is the current of vector mesons with isospin $I=1$.

The dispersion relation is given by
\be
\Pi^{(\rho)} (Q^2) = \frac{1}{12 \pi^2} \int\limits^{\infty}_{4
m^2_{\pi}}~ \frac{R^{I=1} (s)}{s + Q^2} ds
\ee

For the sake of simplicity it is written without subtractions. It
is shown below that the divergent terms in (3) cancel. In the QCD
model with an infinite  number of narrow resonances with the
masses $M_k$ and and the electronic widths $\Gamma^{ee}_k$,
function $R^{I=1}(s)$ has the form
\be
R^{I=1} (s) = \frac{9 \pi}{\alpha^2}~ \sum\limits^{\infty}_{k=0}~
\Gamma^{ee}_k M_k \delta(s - M^2_k) \ee where $\alpha=1/137$.
Formula (4) obviously contradicts to experiments. Let us  replace
$\delta(s-M_k^2)\to
(1/\pi)M_k\Gamma_k/[~(s-M^2_k)^2+M^2_k\Gamma_k^2~],$ $\Gamma_k$ is
the  total width of $k$-th resonance. Then we get, instead of
Eq.(4) $$R^{I=1}(s) =\frac{9}{\alpha^2} \sum^{\infty}_{k=0}
\frac{\Gamma^2_k
M^2_k}{(s-M^2_k)^2+M^2_k\Gamma^2_k}\eqno{\mbox{(4a)}}$$ If the
total widths of all resonances $\Gamma_k$ are much smaller than
their masses $M_k$, the results of the integration of (4) and (4a)
with smooth functions coincide. If $M_k\Gamma_k \gg
M^2_k-M^2_{k-1}$ for $k >3$   function (4a) is described by a
smooth curve for $s
> M^2_3$.   When fulfilling  these conditions  formula (4a)
is consistent with experimental data of $R^{I=1}(s)$. Formula (4)
will be used only under the integral with a smooth function.

Using (4), recast (3) into the form
\be
\Pi^{(\rho)} (Q^2) = \frac{3}{4 \pi \alpha^2}~
\sum\limits^{\infty}_{k=0} ~ \frac{\Gamma^{ee}_k M_k}{s_k + Q^2}
\ee

The polarization operator (5) can be rearranged into a form with
the separated unit operator. The remainder of the polarization
operator can be associated with gluon condensate and with
contribution of higher dimensional operators. To do this we
transform the sum in (5) into an integral by means of the
Euler-Maclaurin formula [10] beginning from $k=1$. We have
$$ \Pi^{(\rho)}(Q^2) = \frac{3}{4 \pi \alpha^2} \left
\{\int\limits^{\infty}_{(m_u+m_d)^2}~ \frac{\Gamma^{ee}_k
M_k}{s_k+Q^2}~ \frac{dk}{ds_k} ds_k - \int
\limits^{s_1}_{(m_u+m_d)^2} ~ \frac{\Gamma_k^{ee} M_k}{s_k+Q^2}~
\frac{dk}{ds_k} ds_k+ \frac{\Gamma^{ee}_0 M_0}{s_0+Q^2} + \right .
$$
\be
\left . \frac{1}{2}~ \frac{\Gamma^{ee}_1 M_1}{s_1+Q^2} -
\frac{1}{12}~ \frac{d}{dk} \Biggl (\frac{\Gamma^{ee}_k
M_k}{s_k+Q^2} \Biggr ) \Biggl \vert_{k=1} + \frac{1}{720} ~
\frac{d^3}{dk^3} \Biggl (\frac{\Gamma^{ee}_k M_k}{s_k+Q^2} \Biggr
) \Biggr \vert_{k=1} - ... \right \} \ee

The operator expansion for $\Pi^{(\rho)}$, which is valid at high
$Q^2$ has the form [11]
\be
\Pi^{(\rho)}(Q^2) = \int\limits^{\infty}_{(m_u+m_d)^2} ~
\frac{R^{I=1}_{PT}(s) ds}{s+Q^2} + C_2/Q^2 + C_4/Q^4 + C_6/Q^6 +
... \ee
\be
C_2 = 0 \ee
\be
C_4 = \frac{1}{24} \langle 0 \vert \frac{\alpha_s}{\pi} G^2 \vert
0 \rangle + \frac{1}{2} (m_u \langle 0 \vert \bar{u} u \vert 0
\rangle + m_d \langle 0 \vert \bar{d} d \vert 0 \rangle ) \ee
$$ C_6 = -\frac{1}{2} \pi \alpha_s \langle 0 \vert (\bar{u}
\gamma_{\mu} \gamma_5 t^a u - \bar{d} \gamma_{\mu} \gamma_5 t^a
d)^2 \vert 0 \rangle $$
\be
-\frac{1}{9} \pi \alpha_s \langle 0 \vert (\bar{u} \gamma_{\mu}
t^a u + \bar{d} \gamma_{\mu} t^a d) \sum\limits_{q=u,d,s} \bar{q}
\gamma_{\mu} t^a q \vert 0 \rangle . \ee

In the region where the operator expansion is valid,
$\Pi^{(\rho)}(Q^2)$ should not differ significantly from
$\Pi^{(\rho)}_{theor}(Q^2)$ (see eq.(7)]. For this reason we
equate the first term on the right-hand part of (6) to the first
term on the right-hand part of (7).
\be
\frac{3}{4 \pi \alpha^2} \int\limits^{\infty}_{(m_u+m_d)^2}~
\frac{\Gamma^{ee}_k M_k}{s_k+Q^2} \frac{dk}{ds_k} ds_k =
\frac{1}{12 \pi^2}~ \int\limits^{\infty}_{(m_u+m_d)^2}~
\frac{R^{I=1}_{PT}(s)ds}{s + Q^2} \ee

We consider the equality (11) as an ansatz that makes it possible
to separate large terms associated with the unit operator from
small terms associated with condensates and consider it as an
equation for $\Gamma^{ee}_k$. Eq.(11) has one and only one
solution\footnote{The analogous formula in nonrelativistic quantum
mechanics $$\mid\psi_k(0)\mid^2=\frac{m^{3/2}}{\sqrt{2}\pi^2}
E^{1/2}_k\frac{dE_k}{dk}~~~~~~\mbox{(12a)}$$ has accuracy $\la
2\%$ for usual potential.}
\be
\Gamma^{ee}_k = \frac{2 \alpha^2}{9 \pi} R^{(I=1)}_{PT} (s_k)
M^{(1)}_k \ee

This result follows from uniqueness of the theorem for analytic
functions. The solution (12) is obtained by equating the jumps on
the cut in (11). In eq.(12) and in the following formulae we use
the notation
\be
s_k =M^2_k,~~~M^{(l)}_k \equiv d^l M_k/dk^l, ~~ s^{(l)}_k \equiv
d^l s_k/dk^l \ee

It can be proved that the function $s_k \equiv s(k)$ specified at
the points $k = 0,1,2, ...$ can be extended to an analytic
function of the complex variable $k$ with a cut along the negative
axis [3]. We will not employ the analytic properties of the
function $s(k)$. It is assumed that the function $s(k)$ is
continuous, differentiable with respect to $k$ at $k=1$. The
derivatives $s^{(l)}_1$ will be considered as parameters.

The QCD model with an infinite number of vector mesons was used in
[5,12] to calculate the contribution of strong interaction to
anomalous magnetic moment of muon. The result was found to be
\be
a^{hadr}_{\mu} = ((g-2)/2)_{hadr} = \frac{\alpha}{3 \pi^2}
\int\limits^{\infty}_{4 m^2 _{\pi}} ds K(s) R(s)/s = 678(7) \cdot
10^{-10} \ee

The result given by eq.(14) should be compared with the recent,
more precise, $a^{hadr}_{\mu}$ value calculated by integrating
formula (14) with the cross sections measured from annihilation
$e^+e^-$ to hadrons [13-15]
$$ a^{hadr}_{\mu} = 6847(70) \cdot 10^{-11} ~~~ [13] $$ $$
a^{hadr}_{\mu} = 6831(61) \cdot 10^{-11} ~~~ [14] $$
\be
a^{hadr}_{\mu} = 6909(64) \cdot 10^{-11} ~~~ [15]\ee In addition,
the QCD model with an infinite number of vector mesons was used in
[4,5,12] to calculate the contribution of strong interaction to
quantity $\alpha(M^2_z)$. It was found that
\be
\delta \alpha_{hadr} = \frac{\alpha M^2_z}{3 \pi} P
\int\limits^{\infty}_{4 m^2_{\pi}}~ \frac{R(s) ds}{(M^2_z-s)s} =
0.02786(6) \ee

This result should be compared with the results $\delta
\alpha_{hadr}$ = 0.02744(36) [16], ~0.02803(65) [17], ~0.0280(7)
[18], ~0.02754(46) [19], ~0.02737(39) [20], ~0.02784(22) [21], ~
0.02778(16) [22], ~0.02779(20) [23], ~0.02770(15) [24], ~0.02787
(32) [25], ~0.02778(24) [26], ~0.02741 (19) [27], 0.02763(36)
[28], and 0.02747(12) [29] which were obtained by calculating the
integral (16) with the experimental cross section from $e^+e^-$
into hadrons. We emphasize that the quantity $\delta
\alpha_{hadr}$ is calculated in (16) with the highest accuracy.

It is important to note that the integrals which describe hadronic
contributions to the $(g-2)$ factor for muon and to
$\alpha(M^2_z)$ are determined by different regions: the integrals
for hadronic contributions to the $(g-2)$ factor is governed by
the region of small $s \sim m^2_{\rho}$, while the integral for
hadronic contribution to $\alpha(M^2_z)$ is dominated by large $s
\sim M^2_z$. From the above that accuracy of calculations by MINVM
is about 1\%. This accuracy is sufficient for calculations of
gluon and four-quark condensates. It is obviously that MINVM can
be used only under the integral.

\vspace{3mm}

\begin{center}
{\bf III. MAGNITUDE OF GLUON AND FOUR-QUARK CONDENSATES\\
FROM
ANALYSIS OF $\rho$-MESON FAMILY}
\end{center}

At present, three mesons of $\rho$-family with the masses $M_0 =
0.7711 \pm 0.0009 GeV$, ~ $M_1 = 1.465 \pm 0.025 GeV$, ~ $M_2 =
1.7 \pm 0.02 GeV$ have been found. The electronic width of
$\rho(770)$ is $\Gamma^{ee}_0 = (6.85 \pm 0.11) keV$ [30].

Comparing (6) and (7) at high $Q^2$ and using formula (12), we
arrive at
$$ C_2 = \frac{1}{12 \pi^2} \left \{ -
\int\limits^{s_1}_{(m_u+m_d)^2} R^{I=1}_{PT} (s) ds + R^{I=1}_{PT}
(s_0) s^{(1)}_0 + \right . $$
\be
\left .\frac{1}{2} R^{I=1}_{PT} (s_1) s^{(1)}_1 - \frac{1}{12}
[R^{I=1}_{PT} (s_1) s^{(1)}_1 ] ^{(1)} \right \} \ee

$$ C_4 = \frac{1}{12 \pi^2} \left \{
\int\limits^{s_1}_{(m_u+m_d)^2}~ R^{I=1}_{PT} (s) s ds -
R^{I=1}_{PT} (s_0) s_0 s^{(1)}_0 - \frac{1}{2} R^{I=1}_{PT} (s_1)
s_1 s^{(1)}_1 + \right . $$
\be
\left . \frac{1}{12} [R^{I=1}_{PT} (s_1) s_1 s^{(1)}_1 ] ^{(1)}
\right \} \ee
$$ C_6 = \frac{1}{12 \pi^2} \left \{ -
\int\limits^{s_1}_{(m_u+m_d)^2}~ R^{I=1}_{PT} (s) s^2 ds +
R^{I=1}_{PT} (s_0) s^2_0 s^{(1)}_0 + \frac{1}{2} R^{I=1}_{PT}
(s_1) s^2_1 s^{(1)}_1 - \right . $$
\be
\left . \frac{1}{12} [R^{I=1}_{PT} (s_1) s^2_1 s^{(1)}_1 ]^{(1)}
\right \} \ee

We discard the terms with small coefficients 1/720, 1/30240 in
formulae (17)-(19). Let us write $R^{I=1}_{PT}$ in the form
\be
R^{I=1}_{PT} (s) = \frac{3}{2} (1 + r(s))
\ee

Function $r(s)$ is calculated in [6]. Neglecting the terms
associated with $u$- and $d$-quark masses we get
\be
C_2=\frac{1}{8\pi^2}\Biggl\{ -s_1 +s_0^{(1)}+\frac{1}{2}
s_1^{(1)}-\frac{1}{12}s^{(2)}_1 -\int\limits^{s_1}_0 r
(s)ds+r(s_0)s^{(1)}_0
+\frac{1}{2}r(s_1)s^{(1)}_1-\frac{1}{12}[r(s_1)s^{(1)}_1]^{(1)}\Biggr\}=0\ee
 $$C_4 =\frac{1}{8\pi^2}\left \{\frac{1}{2}s^2_1 - s_0 s^{(1)}_0
-\frac{1}{2}s_1 s^{(1)}_1 +\frac{1}{12} (s_1
s^{(1)}_1)^{(1)}+\int\limits^{s_1}_0 s r
(s)ds-s_0s_0^{(1)}r(s_0)-\right.$$
\be
\left.-\frac{1}{2}s_1s_1^{(1)}r(s_1)+\frac{1}{12} [s_1s_1^{(1)}
r(s_1)]^{(1)}\right\}\ee
$$ C_6 =\frac{1}{8\pi^2}\left\{- \frac{1}{3}s^3_1 +s^2_0 s^{(1)}_0
+\frac{1}{2} s^2_1 s^{(1)}_1 -\frac{1}{12} (s^2_1 s^{(1)}_1)^{(1)}
-\int \limits^{s_1}_0 s^2 r(s)ds +\right.$$
\be
\left.+ s^2_0 s^{(1)}_0 r(s_0) +\frac{1}{2}s^2_1 s^{(1)}_1 r(s_1)
- \frac{1}{12} (s^2_1s^{(1)}_1 r(s_1))^{(1)} \right\}\ee

Using (21) to eliminate the unobservable  quantity $s^{(2)}_1$ we
reduce (22) and (23) to the form
$$\langle 0\mid \frac{\alpha_s}{\pi} G^2\mid 0 \rangle
=\frac{3}{\pi^2} \left\{- \frac{1}{2} s^2_1 +(s_1-s_0)s^{(1)}_0
+\frac{1}{12} s^{(1)^2}_1 -\int\limits^{s_1}_0(s_1-s)r(s)ds
\right.$$
\be
\left. +(s_1-s_0)s_0^{(1)} r(s_0)
+\frac{1}{12}s^{(1)^2}_1 r(s_1)\right \}\ee
$$C_6 =\frac{1}{8\pi^2} \left\{ \frac{2}{3} s^3_1 -(s^2_1
-s^2_0)s^{(1)}_0 - \frac{1}{6} s_1s_1^{(1)^2}
+\int\limits^{s_1}_0(s^2_1-s^2)r(s)ds -\right.$$ \be \left.
-(s^2_1 - s^2_0)s^{(1)}_0 r (s_0) -\frac{1}{6} s_1 s^{(1)^2}_1
r(s_1)\right\} \ee

Let us calculate $s^{(1)}_0$ and $s^{(1)}_1$.

We obtain from (12) and (20)
\be
s^{(1)}_k =\frac{6\pi \Gamma^{ee}_kM_k}{\alpha^2(1+r(s_k))}\ee

We know only $\Gamma^{ee}_0$, therefore
$$s^{(1)}_0 = (1.554\pm
0.024)~GeV^2~~~~(\Lambda_3=1.565~GeV,~~~r(s_0)=0.204) \eqno(27a)$$
$$s^{(1)}_0 =(1.634 \pm
0.026)~GeV^2~~~~(\Lambda_3=0.618~GeV,~~~r(s_0)=0.144) \eqno(27b)$$

The quantity $s^{(1)}_1$ is determined from trivial equations
$$ s_2 = s_1 + s^{(1)}_1 +\frac{1}{2}s^{(2)}_1 + \frac{1}{6}
s^{(3)}_1 \eqno(28)$$
$$ s_0 =s_1 -s^{(1)}_1
+\frac{1}{2}s^{(2)}_1-\frac{1}{6}s^{(3)}_1\eqno(29)$$

From (28) and (29) we obtain
$$ s^{(1)}_1 = \frac{s_2-s_0}{2} -\frac{1}{6} s^{(3)}_1\eqno(30)$$
and
$$ s^{(2)}_1 =s_0 +s_2 -2s_1 =-0.81 \pm 0.16~GeV^2\eqno(31)$$ To
estimate the last term in (30), we note that $\vert
s^{(2)}_1\vert$ is smaller than $s^{(1)}_1$ (see 32)). We put
$\vert s^{(3)}_1 \vert = \vert s^{(2)}_1 \vert$ and include the
last term in (30) into the error in $s^{(1)}_1$ and obtain finally
$$ s^{(1)}_1 =\frac{s_2-s_0}{2} =1.148 \pm 0.139~GeV^2\eqno(32)$$

Using eqs.(24) and (25) and the values of $s^{(1)}_0$ and
$s^{(1)}_1$ from (27a, 27b, 32), we find that the analysis of the
$\rho$-meson family leads in the 3-loop approximation to the
following results for gluon and four-quark condensates
$$\langle 0 \mid \frac{\alpha_s}{\pi} G^2 \mid 0 \rangle = (0.0744
\pm 0.0227)GeV^4, ~~~ (\Lambda_3 =1.565~GeV) \eqno(33a)$$
$$\langle 0 \mid \frac{\alpha_s}{\pi} G^2 \mid 0 \rangle = (0.112
\pm 0.021)GeV^4, ~~~ (\Lambda_3 =0.618~GeV) \eqno(33b)$$
$$C_6=-0.0072 \pm 0.0041GeV^6, ~~~~~ (\Lambda_3 =1.565~GeV)
\eqno(34a)$$
$$C_6=-0.0115 \pm 0.0034GeV^6, ~~~~~ (\Lambda_3 =0.618~GeV)
\eqno(34b)$$

The errors presented in (33a - 34b) were found from the formulae
$$ \Delta \langle 0\mid \frac{\alpha_s}{\pi} G^2 \mid 0 \rangle
=\frac{3}{\pi^2} \left\{ \Biggl [(s_1-s_0)(1+r(s_0)) \Delta
s^{(1)}_0\Biggr ]^2 + \right.$$ $$\left. +\Biggl [\Biggl (
s^{(1)}_0 -s_1 -\int\limits^{s_1}_0 r(s) ds +s^{(1)}_0
r(s_0)\Biggr )\Delta s_1 \Biggr ]^2 +\Biggl [\frac{1}{6} s^{(1)}_1
(1+r(s_1))\Delta s^{(1)}_1 \Biggr ]^2 \right\}^{1/2}\eqno(35)$$
$$\Delta C_6=\frac{1}{8\pi^2}\left\{\Biggl [ \Biggl ( 2s^2_1 -2s_1
s_0^{(1)} -\frac{1}{6} s^{(1)^2}_1 +2s_1
\int\limits^{s_1}_0r(s)ds\Biggr )\Delta s_1\Biggr ]^2 +\right.$$
$$+ \left.\Biggl [(s^2_1-s^2_0)(1+r(s_0))\Delta s_0^{(1)}\Biggr]^2
+\Biggl [\frac{1}{3}s_1s_1^{(1)}(1+r (s_1))\Delta s_1^{(1)}\Biggr
]^2\right \}^{1/2} \eqno(36)$$

The symbol $\Delta$ in equations in (35) and (36) denotes the
error in the corresponding quality (for example, $\Delta s^{(1)}_1
= 0.139 GeV^2$). The small error connected with the error in
$\Lambda_3$ is taken into account at numerical calculations.

\vspace{3mm}

\begin{center}{\bf IV. MAGNITUDE OF GLUON CONDENSATE FROM THE ANALYSIS\\
OF THE $\omega$-MESON FAMILY}
\end{center}

The polarization operator associated with the isoscalar current
$j^{I=0}_{\mu}(x)$ of light quarks can be written as
$$ i\int\limits d^4 x e^{iqx}\langle 0\mid T \left\{ j^{I=0}_{\mu}
(x) j^{I=0}_{\nu} (0) \right\} \mid 0 \rangle =(q_{\mu}q_{\nu} -
q^2 g_{\mu\nu} )\Pi^{(w)} (Q^2)\eqno(37)$$

where
$$ j^{I=0}_{\mu} (x) =\frac{1}{6} \{ \overline{u}(x)\gamma_{\mu}
u(x)+\overline{d}(x)\gamma_{\mu}d(x)\}\eqno(38)$$

With the exception of (12), all the equations presented above for
the $\rho$-meson family remain in force. The equation that takes
place in (12) for the $\omega$-meson family is
$$ \Gamma^{ee}_k =\frac{1}{9} \frac{\alpha^2}{3\pi} (1+r(s_k))
M^{(1)}_k\eqno(39)$$

Three $\omega$-mesons -- the $\omega$-meson with the mass $M_0 =
0.78257 \pm 0.00012 GeV$ and the electronic width $\Gamma^{ee}_0 =
0.60 \pm 0.02 keV$, the $\omega^{\prime}$-meson with the mass $M_1
= 1.419 \pm 0.031 GeV$, and $\omega^{\prime \prime}$-meson with
the mass $M_2 = 1.649 \pm 0.024 GeV$ [30] -- have been discovered
thus far. Proceeding in the same way as for $\rho$-meson family
and taking into account (39), we obtain
 $$ s^{(1)}_0 = (1.244 \pm 0.041) ~GeV^2, ~~~\Lambda_3 =1.565 GeV
\eqno(40a)$$
 $$ s^{(1)}_0 =( 1.308 \pm 0.044) ~GeV^2, ~~~\Lambda_3 =0.618 GeV
\eqno(40b)$$
$$
 s^{(1)}_1 = (1.053 \pm 0.123) ~GeV^2\eqno(40)$$

Using equations (24) and (35) and the values in (40a,40b,40) we
obtain
$$\langle 0\mid (\alpha_s/\pi)G^2 \mid 0 \rangle = (-0.076 \pm
0.033)~GeV^4 ~~~(\Lambda_3 =1.565 ~GeV)\eqno(41a)$$
$$\langle 0\mid (\alpha_s/\pi)G^2 \mid 0 \rangle = (-0.043 \pm
0.031)~GeV^4 ~~~(\Lambda_3 =0.618 ~GeV) \eqno(41b)$$
$$ C_6 = (0.0079 \pm 0.0052)GeV^6 ~~~(\Lambda_3=1.565~GeV)
\eqno(42a)$$
$$ C_6 = (0.0042 \pm 0.0045)GeV^6, ~~~(\Lambda_3=0.618~GeV)
\eqno(42b)$$

The magnitude of the gluon condensate obtained from different
processes must be equal. We found that the magnitude of the gluon
condensate determined from the analysis of the $\omega$-meson
family is inconsistent with the magnitude of the gluon condensate
from the analysis of the $\rho$-family. The way out of this
situation is proposed in the next sections.

\newpage

\begin{center}
{\bf V. ~$\rho-\omega$-INTERFERENCE AND RESOLUTION\\ OF THE
CONTRADICTION}
\end{center}

Equations (37)-(39) are valid for isospin-zero vector mesons.
However, there is a noticeable isospin-1 admixture in the real
$\omega$-meson. This is obvious, for example, from the fact  that
the width $(\Gamma(\omega \to 2 \pi)$ is nonzero, $$ \Gamma(\omega
\to 2 \pi) = 0.143 \pm 0.024 MeV ~~ [30]$$

Moreover, the ratio of the electronic widths of $\rho$ and
$\omega$ mesons is $(\Gamma^{ee}_{\rho}/\Gamma^{ee}_{\omega} =
11.42(1 \pm 0.037)$, instead of the expected value 9. Let us
represent the $\omega$-meson and $\rho$ states as
$$ \vert \omega\rangle =\mid \omega_0 \rangle +\lambda \mid \rho_0
\rangle\eqno(43)$$ $$\mid \rho \rangle =\mid \rho_0 \rangle
-\lambda \mid \omega_0 \rangle$$ where $\mid \omega_0 \rangle $
corresponds to the I= 0 state and $\mid\rho_0 \rangle$ corresponds
to the I= 1 state. The mixing parameter $\lambda$ can be found in
two ways

1) It follows from the formula $\langle e^+e^-\mid \rho_0 \rangle
=3\langle e^+e^-\mid \omega_0 \rangle$ and the formulae (43) that:
$$ \langle e^+e^-\mid \omega \rangle =(1+3\lambda)\langle
e^+e^-\mid \omega_0 \rangle\eqno(44)$$ $$ \langle e^+e^-\mid \rho
\rangle =(3-\lambda)\langle e^+e^-\mid \omega_0 \rangle\eqno(45)$$
Division of eq.(45) by eq.(44) gives
$$ \frac{3-\lambda}{1+3\lambda} = \frac{\langle e^+e^-\mid \omega
\rangle}{\langle e^+e^-\mid \rho
\rangle}=\sqrt{\frac{\Gamma^{ee}_{\rho}}{\Gamma^{ee}_{\omega}}}=3.379\pm
0.063 \eqno(46)$$ It follows from Eq.(46), that
$$\lambda=-0.034\pm 0.005\eqno(47)$$ It follows Eqs.(43), that
$$ \langle \pi^+\pi^-\mid\omega \rangle =\lambda \langle
\pi^+\pi^-\mid\rho_0 \rangle$$ $$ \langle \pi^+\pi^-\mid\rho
\rangle = \langle \pi^+\pi^-\mid\rho_0 \rangle\eqno(48)$$ The
value $\lambda$ following from Eq.(48) is
$$ \lambda=\frac{\langle \pi^+\pi^-\mid\omega \rangle }{\langle
\pi^+\pi^-\mid\rho \rangle}=-\sqrt{\frac{\Gamma(\omega \to
2\pi)}{\Gamma(\rho\to 2\pi)}}=-0.031\pm 0.0026\eqno(49)$$ The
values $\Gamma(\omega\to 2\pi)$ and $\Gamma(\rho\to 2\pi)$ are
taken Particle Data [30]. In the narrow -- resonance approximation
the parameter $\lambda$ must be real. The values of $\lambda$ from
Eqs.(47) and (49) are in good agreement. Finally we obtain the
averaged mixing parameter $\overline{\lambda}$
$$ \overline{\lambda}=0.0316\pm 0.0023\eqno(50)$$ It follows from
Eq.(44) that $$
\Gamma^{ee}_{\omega_0}=\frac{\Gamma^{ee}}{1+6\lambda}=(0.741\pm
0.027)~keV \eqno(51)$$ Instead of the values $s^{(1)}_0$ from
Eqs.(40a),(40b) we get the new values
$$ s_0^{(1)}=(1.535 \pm
0.057)~GeV^2,~~~(\Lambda_3=1.565~GeV)\eqno(52a)$$
$$s^{(1)}_0= (1.596\pm
0.026)~GeV^2,~~~(\Lambda_3=0.618~GeV)\eqno(52b)$$  corresponding
to the contribution of the isospin $I=0$ in the $\omega$-meson.

The magnitude of gluon condensate following from  the analysis of
the $\omega$-family is
$$\langle 0\mid \frac{\alpha_s}{\pi}G^2\mid 0 \rangle = (0.073 \pm
0.034)~GeV^4,~~~(\Lambda_3=1.565~GeV)\eqno(53a)$$
$$\langle 0\mid \frac{\alpha_s}{\pi}G^2\mid 0 \rangle = (0.108 \pm
0.033)~GeV^4,~~~(\Lambda_3=0.618~GeV)\eqno(53b)$$

This $GC$ magnitude is consistent with the value which follows
from the analysis of the $\rho$-meson family.

For $C_6$ we have
$$ C_6=(-0.0084 \pm
0.0050)~GeV^6,~~~(\Lambda_3=1.565~GeV)\eqno(54a)$$
$$ C_6=(-0.0123 \pm
0.0044)~GeV^6,~~~(\Lambda_3=0.618~GeV)\eqno(55b)$$ The results of
the  calculations of gluon and 4-quarks condensates in 0--3 loops
approximation from the analysis of $\omega$-meson family are
presented in the Tables I,II

\vspace{5mm}

TABLE I. ~~The results of the calculations of the gluon and
4-quarks condensates

\hspace{2.1cm} in the 0--3 loops approximation for
$\Lambda^{new}_3$ from  analysis of

\hspace{2.1cm}  $\omega$-meson family taking into account the
$\rho-\omega$ interference.

\begin{center}

\begin{tabular}{|l|l|l|l|l|l|}\hline
& $\Lambda^{new}_3/GeV$ & $\frac{\langle 0\mid(\alpha_s/\pi)
G^2\mid \rangle}{GeV^4}$ & ~~~$ C_6/GeV^6$& ~~$r(m^2_{\omega}$)&
~~$r(m^{'2}_{\omega'}$)\\ \hline 0 loops & & 0.198$\pm$0.100 &
-0.0218$\pm 0.0034$ & ~~~~~0 & ~~~~~0 \\ \hline 1 loop &
0.618$\pm$ 0.059 & 0.070$\pm $0.034& -0.0082$\pm 0.0050$&
0.201$\pm$0.008 & 0.153$\pm$0.007
\\ \hline 2 loops & 1.192$\pm$0.136 & 0.072$\pm $ 0.034 & -0.0083$\pm$
0.005 & 0.203$\pm$0.008 & 0.161$\pm$0.008 \\ \hline 3 loops &
1.565$\pm$0.193 & 0.073$\pm$0.034 & -0.0084$\pm$0.0050 &
0.202$\pm$ 0.008 & 0.164$\pm$0.008
\\ \hline

\end{tabular}

\end{center}

\vspace{1cm}

TABLE II. ~~The results of the calculations of the gluon and
4-quarks condensates

\hspace{2.1cm} in the 0--3 loops approximation for
$\Lambda^{conv}_3$ from the analysis of  $\omega$-meson

\hspace{2.1cm} family taking into account the $\rho-\omega$
interference.

\vspace{3mm}

\begin{center}

\begin{tabular}{|l|l|l|l|l|l|}\hline
& $\Lambda^{conv}_3/GeV$ & $\frac{\langle 0\mid(\alpha_s/\pi)
G^2\mid \rangle}{GeV^4}$ & ~~~$ C_6/GeV^6$ & ~~$r(m^2_{\omega}$) &
$r(m^{'2}_{\omega'})$\\ \hline 0 loops & & 0.198$\pm$0.100 &
-0.0218$\pm$0.0034 & ~~~~~0 & ~~~~~0
\\ \hline 1 loops & 0.370$\pm$0.019 & 0.097$\pm$0.033
& -0.0111$\pm$0.0046 & 0.159$\pm$0.004& 0.122$\pm$0.003 \\ \hline
2 loops & 0.539$\pm$0.025 & 0.105$\pm$0.033 & -0.0119$\pm$0.0045 &
0.148$\pm$0.003 & 0.114$\pm$0.002
\\ \hline 3 loops & 0.618$\pm$0.29 & 0.108$\pm$0.033 & -0.0123$\pm$ 0.044 &
 0.143$\pm$0.003 & 0.111$\pm$0.002\\ \hline
\end{tabular}
\end{center}

\newpage

\begin{center}
{\bf  VI. The magnitude of gluon and four-quark condensates

from $\rho$-meson family with the account of $\rho -
\omega$-interference}

\end{center}

Because of $\lambda \not= 0$ there is a small admixture of isospin
$I = 0$ into $\rho(770)$ meson. The electronic width of
$\rho(770)$-meson due to isospin $I=1$ is equal to

$$ \Gamma^{ee}_{\rho_0} = \frac{\Gamma^{ee}_{\rho}}{(1 -
\frac{\lambda}{3})^2} = 6.71 \pm 0.11 keV \eqno(56)$$

 The corrected value $s^{(1)}_0$
is equal to

$$ s^{(1)}_0 = (1.521 \pm 0.025) GeV^2, ~~ (\Lambda_3 = 1.565 GeV)
\eqno(57a) $$

$$ s^{(1)}_0 = (1.596 \pm 0.026) GeV^2, ~~ (\Lambda_3 = 0.618 GeV)
\eqno(57b)$$

The corresponding GC magnitude is $$ \langle 0 \vert
\frac{\alpha_s}{\pi} G^2 \vert 0 \rangle = (0.056 \pm 0.023) GeV^4
~~ (\Lambda_3 = 0.565 GeV) \eqno(58a)$$

$$ \langle 0 \vert \frac{\alpha_s}{\pi} G^2 \vert 0 \rangle =
(0.096 \pm 0.021) GeV^4, ~~ (\Lambda_3 = 0.618 GeV) \eqno(58b)$$

For $C_6$ we have $$ C_6 = (-0.0051 \pm 0.0045) GeV^6, ~~
(\Lambda_3 = 1.565 GeV) \eqno(59a)$$

$$ C_6 = (-0.0097 \pm 0.0039)GeV^6, ~~ (\Lambda_3 = 0.618 GeV)
\eqno(59b) $$ The results of the calculations of gluon and
4-quarks condensates in 0--3 loops approximation from analysis of
the $\rho$-meson family are presented in the Tables III, IV.

\vspace{5mm}

TABLE III. ~~The results of the calculations of the gluon and
4-quarks condensates

\hspace{2.1cm} in the 0--3 loops approximation for
$\Lambda^{new}_3$ from  analysis of

\hspace{2.1cm}  $\rho$-meson family taking into account the
$\rho-\omega$ interference.

\begin{center}

\begin{tabular}{|l|l|l|l|l|l|}\hline
& $\Lambda^{new}_3/GeV$ & $\frac{\langle 0\mid(\alpha_s/\pi)
G^2\mid \rangle}{GeV^4}$ & ~~~$ C_6/GeV^6$& ~~$r(m^2_{\rho}$)&
~~$r(m^{2}_{\rho'}$)\\ \hline 0 loops && 0.197$\pm$0.059 &
-0.0211$\pm$0.0025 & ~~~~0 & ~~~~0 \\ \hline 1 loop &
0.618$\pm$0.059 & 0.049$\pm$0.023 & -0.0044$\pm$0.0045 &
0.202$\pm$ 0.008 & 0.151$\pm$0.007 \\ \hline 2 loops &
1.192$\pm$0.136 & 0.055$\pm$0.023 & -0.0050$\pm$0.0045 &
0.204$\pm$0.008 & 0.159$\pm$0.008 \\ \hline   3 loops &
1.565$\pm$0.193 & 0.056$\pm$0.023 & -0.0051$\pm$0.0045 &
0.203$\pm$0.008 & 0.162$\pm$0.008 \\ \hline
\end{tabular}
\end{center}

\newpage

TABLE IV. ~~The results of the calculations of the gluon and
4-quarks condensates

\hspace{2.1cm} in the 0--3 loops approximation for
$\Lambda^{conv}_3$ from  analysis of

\hspace{2.1cm}  $\rho$-meson family taking into account the
$\rho-\omega$ interference.

\begin{center}

\begin{tabular}{|l|l|l|l|l|l|}\hline
& $\Lambda^{new}_3/GeV$ & $\frac{\langle 0\mid(\alpha_s/\pi)
G^2\mid \rangle}{GeV^4}$ & ~~~$ C_6/GeV^6$& ~~$r(m^2_{\rho}$)&
~~$r(m^{2}_{\rho'}$)\\ \hline 0 loops && 0.197$\pm$0.059 &
-0.0211$\pm$0.0025 & ~~~~0 & ~~~~0 \\ \hline 1 loop &
0.370$\pm$0.019 & 0.070$\pm$0.034 & -0.0082$\pm$0.0050 &
0.159$\pm$ 0.006 & 0.151$\pm$0.007 \\ \hline 2 loops &
1.539$\pm$0.025 & 0.093$\pm$0.022 & -0.0094$\pm$0.0039 &
0.149$\pm$0.003 & 0.113$\pm$0.002 \\ \hline   3 loops &
1.618$\pm$0.029 & 0.096$\pm$0.021 & -0.097$\pm$0.039 &
0.144$\pm$0.003 & 0.109$\pm$0.002 \\ \hline

\end{tabular}
\end{center}

Since $\rho(770)$ has a small admixture of isospin $I=0$, the
decay $\rho \to \pi^+\pi^-\pi^0$ exists with the width

$$ \Gamma_{\rho\to\pi^+\pi^-\pi^0} = \lambda^2 \Gamma_{\omega \to
\pi^+\pi^-\pi_0} = (7.23 \pm 1.20) keV \eqno(60) $$

The value

$$ \Gamma_{\rho\to\pi^+\pi^-\pi^0}/\Gamma_{tot} = (4.8 \pm 0.8)
\cdot 10^{-5} \eqno(61) $$

is smaller than the experimental restriction [30]

$$ (\Gamma_{\rho \to \pi^+\pi^-\pi^0}/\Gamma_{tot})_{Exp} <
1.2\cdot 10^{-4} \eqno(62) $$

\vspace{7mm}

\centerline{\bf VII.~ CALCULATION OF THE ELECTRONIC WIDTH $\rho
(1450$).}

\vspace{3mm}

From eq.(26) we have
$$\Gamma^{ee}_1 =\frac{\alpha^2}{6\pi}
\frac{(1+r(s_1))}{M_1}s^{(1)}_1\eqno(63)$$

and using $s^{(1)}_1$ we get from (32)
$$\Gamma^{ee}_1=(2.57\pm
0.31)~keV,~~~(\Lambda_3=1.565~GeV)\eqno(64a)$$
$$\Gamma^{ee}_1=(2.46\pm
0.30)~keV,~~~(\Lambda_3=0.618~GeV)\eqno(64b)$$ The results
presented in (54a,54b) are consistent with the value
$\Gamma^{ee}_1 = (2.5 \pm 0.9) keV$ obtained in [31].

\vspace{7mm}

\centerline{\bf VIII.~ CALCULATION OF THE ELECTRONIC WIDTH
$\omega(1420)$}

\vspace{3mm}

We have from eq.(39)
$$ \Gamma^{ee}_1
=\frac{\alpha^2}{54\pi}~\frac{(1+r(s_1)}{M_1}s^{(1)}_1\eqno(65)$$
and using $s^{(1)}_1$ from (40) we obtain
$$\Gamma^{ee}_1=(0.27\pm0.03)~keV,~~~(\Lambda_3=1.565~GeV)\eqno(66a)$$
$$\Gamma^{ee}_1=(0.26\pm0.03)~keV,~~~(\Lambda_3=0.618~GeV)\eqno(66b)$$

Within the errors, the electronic width of the
$\omega(1420)$-meson is one-ninth  of the electronic width of the
$\rho(1450)$ and is inconsistent with the value $\Gamma^{ee}_1 =
0.15 \pm 0.4) keV$ [31].

\vspace{7mm}

\centerline{\bf IX. ~ CONCLUSION}

\vspace{3mm}

In summary, we present the results of the averaging over the
$\rho$ and $\omega$-families   of the gluon and 4-quark
condensates in the 0-3 loop approximation in Table V,VI.

\vspace{7mm}

TABLE V. ~~The averaged results of the calculation of the gluon
and 4-quarks

\hspace{2.3cm}condensates in  0--3 loops approximation for
$\Lambda^{new}_3$ from  analysis of

\hspace{2.3cm}  $\rho$ and $\omega$  families taking into account
the $\rho-\omega$ interference.

\begin{center}

\begin{tabular}{|l|l|l|l|}\hline
& $\Lambda^{new}_3/GeV$ & $\langle 0\mid(\alpha_s/\pi) G^2\mid
\rangle /GeV^4$ & ~~~$ C_6/GeV^6$ \\ \hline 0 loops &&
0.198$\pm$0.100 & -0.0218$\pm$0.0034   \\ \hline 1 loop &
0.618$\pm$0.059 & 0.056$\pm$0.019 & -0.0061$\pm$0.0033
\\ \hline 2 loops & 1.192$\pm$0.136 & 0.061$\pm$0.019 &
-0.0065$\pm$0.0034   \\ \hline 3 loops & 1.565$\pm$0.193 &
0.062$\pm$0.019 &-0.0066$\pm$0.0034
\\ \hline

\end{tabular}
\end{center}

\vspace{7mm}

TABLE VI. ~~The averaged results of the calculation of the gluon
and 4-quarks

\hspace{2.3cm}condensates in  0--3 loops approximation for
$\Lambda^{new}_3$ from  analysis of

\hspace{2.3cm}  $\rho$ and $\omega$  families taking into account
the $\rho-\omega$ interference.

\begin{center}

\begin{tabular}{|l|l|l|l|}\hline
& $\Lambda^{new}_3/GeV$ & $\langle 0\mid(\alpha_s/\pi) G^2\mid
\rangle /GeV^4$ & ~~~$ C_6/GeV^6$\\ \hline 0 loops &&
0.198$\pm$0.100 & -0.0218$\pm$0.0034   \\ \hline 1 loop &
0.370$\pm$0.019 & 0.088$\pm$0.018 & -0.0096$\pm$0.0030
\\ \hline 2 loops & 0.539$\pm$0.025 & 0.096$\pm$0.018 &
-0.0105$\pm$0.0029  \\ \hline 3 loops & 1.618$\pm$0.029 &
0.100$\pm$0.018 &-0.0108$\pm$0.0029
\\ \hline

\end{tabular}
\end{center}

It is seen from Tables I-VI that the expansion of the gluon and
4-quark condensates in terms of $\alpha_s$ is very good. The good
convergence in $\alpha_s$ is due to the improved perturbative
theory [52,50,6]  used in the paper.

The magnitude of the gluon condensate from the analysis of
families of $J/\Psi$ and $\Upsilon$ mesons was obtained in paper
[3].
$$ 0.04 \leq \langle (\alpha_s/\pi) G^2 \rangle \leq 0.105
~GeV\eqno(67)$$

The first-order terms in $\alpha_s$ and Coulomb terms of all
orders in $\alpha_s/v_k$ are taken into account when obtaining
(67). The result (67) is in a lightly better agreement with
$\Lambda^{new}_3$ than with $\Lambda^{conv}_3)$.

The conventional magnitude of gluon condensate is the value
obtained by Shifman, Vainstein and Zakharov in their basic paper
[11] from analysis of the $J/\Psi$ family
$$ \langle (\alpha_s/\pi) G^2 \rangle =0.012~GeV^4 \eqno(68)$$

But as was shown in ref.[3] the model used by Shifman et al. in
[11] to describe the experimental function $R_c(s)$ in the form of
the sum of $\delta$-functions (related to the observed resonances
from the $J/\psi$ family) and a plateu contradicts the Wilson
operator expansion in terms which are due to the gluon condensate.
Later, there were many attempts to determine the gluon condensate
by considering various processes within various approaches
[32-51]. But in these works either $R_c$ contradicts the Wilson
operator expansion or $\Lambda_3$ is very small, $\Lambda_3 \sim
100 MeV$.

\vspace{7mm}

\centerline{\bf ACKNOWLEDGEMENTS}

\vspace{3mm}

The author thanks V.A.Novikov for useful discussions.
 The research described in this publication was made
possible in part by Grant No.RP2-2247 of US Civilian Research and
Development Foundation for Independent State of the Former Soviet
Union (CRDF), by the Russian Found of Basic Research, Grant
No.00-02-17808 and INTAS Call 2000, Project No.587.

\end{document}